# A pattern based methodology for evolution management in business process reuse

Hanae Sbai, Mounia Fredj and Laila Kjiri

AlQualsadi team, ENSIAS, Mohammed V Souissi University
Rabat, Morocco

**Abstract**
**Today, there are Process-Aware Information Systems (PAIS) with a set of business process models which vary over time to meet the new requirements. In a competitive environment, the key challenge of enterprises is to reduce the cost and time of process design and application development. For this purpose, research on reuse in business process management have introduced the concept of configurable process models which attempts to manage business process variability, by integrating a set of process variants in a single model. In this context, many research works were interested in creating and elaborating configurable process models. However, this has become insufficient since the configurable process model should itself evolve to add new variations. In turn, this requires a comprehensive support for managing the evolution of configurable process models. In this paper, we present a complete pattern based methodology for managing the evolution of configurable process models in terms of activities, data and resources. Our objective is to propose a process patterns system for guiding designers in modeling and evolving configurable process models. Furthermore, our process patterns system will be used for an automated support so as to manage the evolution of configurable process models.**

*Keywords: PAIS, business process variability, process evolution, configurable process model, process pattern*

## 1. Introduction

BPM (Business Process Management) is an approach which suggests the alignment of information systems with business processes through a process-oriented approach. This increases the adoption of another kind of information systems called "Process Aware Information Systems" (PAIS) [1]. The main goal of BPM is to reduce the cost and time of process design and application development. However, there are process oriented software systems [2] with a set of business process models which usually vary over time to meet the new requirements. Therefore, maintaining process models repositories with a set of process variants becomes too costly, and implementing it separately is too inefficient. In addition, the customization of business process models is done manually, which both time and cost is consuming. In this context, the reference process modeling approach aims to allow reusing business process management by introducing the concept of *Configurable Process Model* (CPM). The main reason behind using this concept of process model *reuse* is to avoid designing process models which have been defined and used by others [3]. A configurable process model represents the concept of a *reference process model* which identifies common practices and activities of organizations. These models are intended to provide reusing variation options which have been integrated within the model beforehand [3]. For example, CPMs of an E-health-care system capture common practices for handling medical examination, and may contain a hundred of process variants with hospital differences. So, the CPM can be customized using possible variations integrated in the model. Existing approaches on *managing business process variability* [3] are motivated by creating configurable process models and extending the business process modeling language to support variability. However, if there is a need to introduce new variation options, the CPM should evolve to support new changes. Much attention has, therefore, to be made on enhancing process model configuration with adaptation mechanisms to add new behavior to the configurable process models. This has motivated the emergence of research works on configurable process evolution. Thus, the concept of *business process evolution* has been widely discussed in the field of BPM [1] [4] [5] [6]. In this context, we mention the work of change patterns [1] which attempts to allow the creation and adaptation of a single process model. These patterns are not sufficient to cope with configurable process models evolution, which is a collection of related process variants. Consequently, this requires the support of the managing evolution of configurable process models.

For this reason, we propose a **methodology for evolving configurable process models**. The proposed approach is based on a system of **process patterns** [7] [8], which capitalizes a set of processes and model solutions that carry out every step of the evolution process. Thus, in order to ensure quality design, an **evolution meta-model has been proposed in** [8]. It allows to describe configurable process model changes when applying the proposed process patterns and enables evolution traceability when implementing the proposed process patterns. In this paper, we present the complete process patterns based approach to design and evolve configurable process models along the business process life cycle through a reuse process.

The remainder of this paper is structured as follows. Section 2 introduces configurable process models and variability concepts upon which this work is based. Section 3 provides an overview of contemporary works on business process variability and evolution. Section 4 is presents our proposals for managing the evolution of configurable process models. Section 5 concludes the paper.

## 2. Background and foundations

In this section, we begin by introducing business process variability which is a prerequisite for understanding the core concept of a configurable process model.

A *configurable process model* deals with how to model business processes that are similar to one another in many ways, yet different in some other ways from one organization, project or industry [9]. A *configurable process model* is a combination of *variability* and business process concepts. This combination allows to describe a set of process variants by extending existing business process modeling languages to support variability.

Variability concepts were first introduced in the field of software product line engineering [10]. Variability is defined as the ability to change or customize a software system [11]. It refers to the diversity of variations of the manufacturing processes for producing product variants in a product family [12].

For a given business process, the variability is concerned by defining which parts of the process may vary over time or within a domain space called « variation point », and what are the different realizations of each variation point called «variant ». For modeling variability, the work of [11] defines types of variation points and dependencies relationships.

There are three types of variation points:
- *Optional*: it corresponds to the choice of selecting zero or one from one or more variants.
- *Alternative*: it corresponds to the choice of selecting only one variant.
- *Optional alternative*: it is an optional variation point with alternative variants.

Concerning the relationships between variants, they can be of two types:
- *Inclusion*: it specifies that the choice of a variable element requires the presence of another variable element.
- *Exclusion*: it specifies that the choice of a variable element excludes the presence of another element.

In the literature, PESOA project [11] distinguishes two variability mechanism categories, namely basic and composite variability mechanisms. The first category is composed of three types of mechanisms:
- *Encapsulation of varying sub-processes*: it allows the insertion of different sub-processes variants hidden by the invariant interface.
- *Parameterization*: behavioral variants are integrated in the process and activated by configuring the process with corresponding parameter values.
- *Variability in data type*: it represents the variations of the data stored in the process.

As for the second category (i.e. composite variability mechanisms) we find:
- *Inheritance*: it allows the replacement or addition of model elements in the derived process diagram.
- *Extension*: it is the insertion of encapsulated optional sub-processes at the extension point which refers to the place where the process can be extended.

In this section, we have dealt with the concept of business process variability and its mechanisms. In what follows, we present a state of the art of reuse in business process management.

## 3. Related work

The aim of reuse in business process management is to avoid designing process models which have already been defined. This has led to the introduction of a *reference* process model that defines the common practices of organizations that can be reused. However, customizing these models is done manually, which is a difficult and time-consuming task. In this context, configurable process models have been developed to ensure a **systematic reuse** of reference process models with managing *business process variability* [9] which is a key challenge in business process reuse.

Solving the problem of business process variability requires a business process variability modeling language, an automated support for configuring business process and a *support for configurable process evolution*. Therefore, research studies that seem to be of utmost importance to our work are those that are undertaken on i) business process variability modeling languages, i)) configuring business processes and iii) business process evolution.

## 3.1 Business process variability modeling languages

Concerning *the configurable node based approach*, [13] proposed the C-EPC (Configurable- Event Process Chain) which is based on configurable nodes variability technique This proposal extends the existing business process modeling language EPC by adding configurable elements for explicitly representing variability. To add configuration opportunities for workflow models, Gottschalk *et al.* introduce the "Configurable Yet A new Workflow Language" (C-YAWL) which is an extended version of the YAWL language [14]. La Rosa proposes C-iEPC (Configurable-Integrated Event Process Chain) language which is an enhancement of C-EPC to support roles and objects configuration [9]. An automatic mechanism to create configurable process models by merging business process is introduced by [15]. For *the annotation based approach,* [11] put forward Variant-Rich process models that are based on extending the concept of UML stereotype to represent variability in BPMN process models, and in particular at the activity/control flow level. A hierarchical method based on extending UML2 profile to represent variability is presented by [16]. This work is interested in control flow, dataflow and action variability. For the *CVL based approach*, Ayora introduces a separate model for modeling variability at the control flow and task related elements level which is based on extending the Common Variability Language (CVL) approach [17](a domain-independent language for specifying and resolving variability over any instance of any MOF-compliant meta-model [18]).

All these approaches extend business process modeling languages with variability techniques that enable the creation of configurable process models.

In the next section, we present works that have dealt with the configuration of business process models.

## 3.2 Support for configuring business process models

Concerning the works on *behavioral model*, Gottschalk proposes a configuration support for C-EPC models and a questionnaire based approach for controlling configuration [3]. An enhancement of this work is introduced by La Rosa *et al.* who propose a multi-perspective approach for configuring configurable process models using C-iEPC language [11]. To preserve configurable process models correctness, Van der Aalst *et al.* suggest a set of syntactic and semantic constraints to ensure sound configuration process variants [20]. In order to support flexible process variants management and retrieval, Lu *et al.* propose two approaches, namely a process constraint based approach and a query formalization approach [21]. Santos *et al.* develop a non-functional based approach for managing business process variability [22]. A mining technique based approach used to create configurable process models is introduced by Buijs *et al.* [23]. All these works have served as a basis for implementing the process configuration in different stages of the process lifecycle. In this context, the first toolset called "SYNERGIA" has been developed by La Rosa *et al.* [24]. It provides a comprehensive support for the configurable process modeling notations (C-EPC, C-iEPC and C-YAWL). In order to manage a large set of process variants, the authors provide the APROMORE tool which brings together a rich set of features for the analysis, management and usage of large sets of process models [25]. For structural model works, Schnieders *et al.* define a set of variability mechanisms and discuss their possible implementations for process family systems in the context of PESOA research project [26]. To design process families, Hallerbach *et al.* develop the PROVOP framework for managing a family of process variants with paying attention to the application context [27]. It has been served as basis of configuration process implementation by developing the PROVOP prototype which is implemented in the ARIS business architect. In addition, Rolland *et al.* propose a business goal based approach for managing business process variants [28]. A functional requirement based approach for preserving process variants correctness is introduced by Groner *et al.* [29].

From the discussion above, it is quite clear that the reviewed works are interested in configuring the collection of process variants integrated into a configurable process model. This becomes insufficient if one day the organization needs to add a new variation point/variant that is not defined in the model beforehand. To meet new needs, the configurable process model should evolve to integrate new variable elements. In the next section, we present works on business process evolution.

## 3.2 Business process evolution

To *evolve a business process model*, change patterns and change support features [1] provides a set of patterns for a single process model adaptation. This work is considered as a support of many implementations. We also quote frameworks such us Adept2 [30] and AristaFlow [31] for dynamic adaptation as well as[32] for ensuring safety of workflow dynamic adaptation Gschwind *et al.* propose a set of patterns based on workflow patterns to support business process modeling [33]. To extend *configurable process model for supporting adaptation techniques*, [34] proposed a detailed definition of a set of generic adaptation concepts for adapting EPC reference process models Specialization of C-EPC configurable process models to add new functions has been ensured by ADOM-EPC formalism [35]. Lately, Ayora *et al.* discuss the problem of dynamic evolution of process families [4] and introduce a pattern based approach to manage evolution of process families in terms of activity [5].

As is shown above, research studies on evolution in BPM, and particularly on change patterns, are interested in managing business process evolution. However, these works are not sufficient to evolve configurable process models because they do not support the variability concept introduced by configurable process models. Furthermore, works on extending configurable process model adaptability do not establish a comprehensive process for guiding configurable process model evolution. Hence, our aim in the present paper is to provide a methodology based on a process patterns system for guiding designers in configurable process model evolution. This guide will provide support for using process patterns that are applicable in the designer context, and giving feedback if the applications of some adaptation scenarios lead to a modeling error. Thus, this support can allow us to trace the sequence of operations applied using process patterns during the evolution process. As our approach is based on the process pattern concept, we present in the next section the P-SIGMA [36] formalism used to define the proposed process patterns system.

## 5. Our proposal for managing configurable process models evolution

The proposed process patterns system provides an effective guidance for evolving configurable process model evolution in terms of activity, resource and data. Each configurable process model elements (activity, resource, and data) may be a variation point with a set of variants. We describe below our process patterns system (cf. Figure 1). It has been defined by analyzing various types of evolution that can be applied to a configurable process model. So, if there is a need to evolve configurable process model occurs, we can have the following basic evolution types for each process element (activity, data or resource): Insertion/substitution/Deletion of a variation point/variant. The process patterns system can be divided into three sub-systems, namely activity process patterns, resource process patterns and data process patterns sub systems.

For the activity process patterns sub-system we define: Activity Insertion (AI) which is refined by Variation Point Activity Insertion (VPAI), and Variant Activity Insertion (VAI). The Activity Substitution (AS) which is refined by Variation Point Substitution (VPAS), and Variant Activity Substitution (VAS). Activity Deletion (AD) which is refined by Variation Point Deletion (VPD), and Variant Activity deletion (VAD). By the same way, we have defined resource and data process patterns sub- systems.

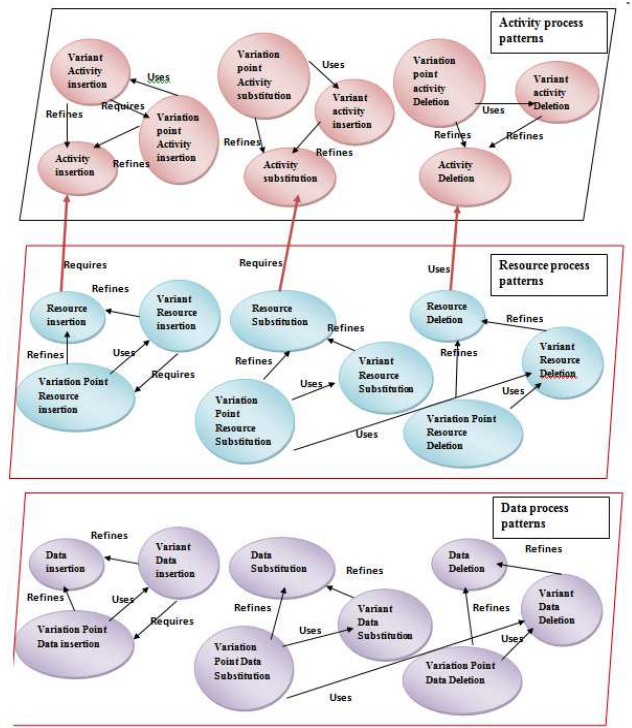

Figure 1: The process patterns system [8]

To guide the designer to evolve a given configurable process model, we define the following evolution constraints:

➢ Insertion of a variation point **requires** the insertion of at least one variant
➢ Insertion of a variant **requires** the presence of a variation point or the transformation of an existing process element to a variation point.
➢ Substitution of a variation point **includes** the substitution of related variants or/and preserving existing variants.
➢ Deletion of a variation point **includes** the deletion of the related variants.
➢ Deletion of a variant **requires** to check the constraints of the other one to be sure that this variant is not required by another variant.

In addition, the proposed process patterns use the following parameters:

| Process patterns parameters | Definitions |
|---|---|
| Type | **Type** of a variation point allows to determine if a variation point is alternative or optional |
| Req_f_VPA | **Req**uired **f**unctionalities of a **V**ariation **P**oint **A**ctivity represent all needed functions of an activity to be performed by a resource. |
| R_f | **R**esource **f**unctionalities represent all functions ensured by the resource to perform an activity. |
| VSC | **V**ariant **S**election **C**ondition represents the condition under which a variant can be selected. |

| | | |
|---|---|---|
| VCC | | Variant Configuration Constraint specifies if the choice of a variant requires or excludes the presence of other variants. |

Table 1 : Process patterns parameters

To describe each process pattern, we use the P-Sigma formalism and in particular, the realization part that gives the solution in terms of *Process Solution* and *Application* of the process Solution. In our case:
- The *Process Solution* is described by an algorithm.
- The *Application* represents the configurable process model solution after applying the process pattern by using the Rich Variant BPMN process model [16].

In the next section, we present some process patterns of the proposed system, namely the Variation Point Activity Insertion and the Variation Point Activity Substitution. We explain in the next section, how to insert a variation point activity by using the VPAI process pattern.

## 5.2 Variation Point Activity Insertion (VPAI) process pattern

When there is a need to add a new variation point activity, a set of process patterns must collaborate to lead this evolution.

| Parts | Fields | Values |
|---|---|---|
| **Interface** | Identification | Variation Point Activity Insertion (VPAI) |
| | Classification | Variant activity, configurable process model |
| | Context | {Variant Activity Insertion (VAI), Data Insertion (DI), Resource Insertion (RI)} |
| | Problem | Insertion of an activity as a variation point in the configurable process model |
| | Force | Allows verifying a set of constraints before inserting a variation point activity. |
| **Realization** | Process Solution | **Needed parameters :**<br>*Req_f_VPA: Required functionalities of a variant activity*<br>*R_f: Resource functionalities*<br>*C_nbr_A: Current number of activities*<br>*Max_nbr_A : Maximum number of activities*<br>*Type_VP: the type of a variation point(it may be optional or alternative)*<br>**Design choices**: (a) add a new variation point or (b) transform an existing activity to a variation point<br>**If** the activity is a **new added variation point then**<br>   Determine the position of the insertion of the **added variation point** in the sequence flow<br>   Insert the type of the added variation point activity // *it can be optional or alternative*<br>   Apply Variant Activity Insertion process pattern // *To insert variant activities*<br>   Insert the required functionalities of the **added variation point activity**<br>   Determine the resource to assign<br>      **If** Rq_f_VPA belongs to f_R **then** assign resource<br>         **If** the assigned resource is a variation point **then**<br>           Apply Variant Resource Insertion process pattern// *the required resource variant must be added*<br>         **Else** transform (Resource to a variation point Resource)<br>         **End if**<br>      **Else** Apply Insertion Resource process pattern // *Insert resource*<br>      **End if**<br>   Insert the flow sequence condition//*If needed insert the condition which to performs the added variation point activity*<br>   Apply the Data Insertion process pattern// *Insert data*<br>**Else if** we transform an existing activity to a variation point **then**<br>   Repeat the same steps as those of the insertion of a new added variation point activity<br>**End if** |
| | Application | **For example, we apply the process solution for (a):**<br>1. We suppose that the designer inserts a new variation point (VP) activity "B" (1)<br>2. We suppose that the designer inserts the variation point activity"B" between "A" and "D ". So:<br>   ✓ The sequence flow {A→D} is deleted (2.1) |

- ✓ The VP activity B is inserted and the sequence flows {A→B} and {B→D} are inserted (2.2)
3. Insert the flow sequence condition "condition= cond1" (3)
4. The designer chooses for "B" the type= "alternative". In this case, the VP activity "B" should be annotated with «VarPoint ». If he chooses type="optional", the activity VP should be annotated "Null".(4)
5. The designer inserts required functionalities of "B": RF1,RF2,RF3 (5)
6. The designer should insert variant activities which depend on "B". In this case, the variant activity insertion process pattern is applied repeatedly to insert variant activities "B1" and "B2". (6)
7. To insert data, the data insertion process pattern is applied to insert in this example "dataobjet (7)
8. The designer chooses the resource R1 to be assigned to the VP activity "B". In this case we can have two scenarios (8):
    - ✓ Scenario1 :{ RF1, RF2, RF3} belong to {FR1, FR2, and FR3}: the resource R1 is assigned. As the resource R1 is a variation point, o the designer should insert a variant resource VR which will be assigned to the variant activities B1 and B2. In this case the variant resource insertion process pattern is applied. (8.1)
    - ✓ Scenario2 :{ RF1, RF2, RF3} do not belong to {FR1, FR2, FR3}. The designer should insert the new resource R2 which is an alternative variation point. In this case, the variation point resource insertion process pattern is applied. (8.2)

In our case, we apply the scenario 1 (8.2)

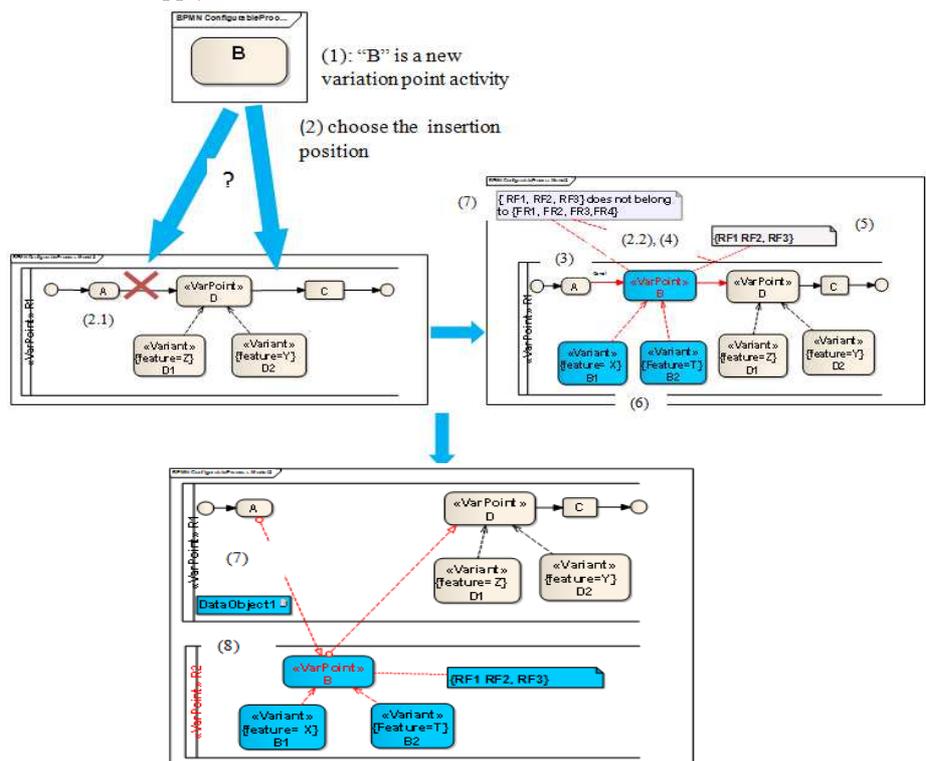

| Relations | Refines | Insertion Activity process pattern (IA) |
|---|---|---|
| | Uses | {Variant Activity Insertion (VAI), Data Insertion (DI), Resource Insertion (RI)} |

Table 2 : The VPAI process pattern

For this process pattern, we obtain the following relations:
- {VPAI} **refines** {AI}: the process pattern VPAI is a specialization of the process pattern AI.
- {VPAI} **uses** {DI}: to insert a new variation point activity we have to insert a related data.
- {VPAI} **uses** {RI}: To insert a new variation point activity, we have to assign a resource.
- {VPAI} **uses** {VAI}: To insert a new variation point activity, we have to insert at least a default variant activity.

In the next section, we present the Variation point Activity Substitution process pattern which provides all steps that we should apply to substitute a variation point activity by another.

## 5.4 Variation Point Activity Substitution (VPAS) process pattern

The application of the variation point activity substitution can invoke substitution of data (variation point/variant), substitution of a resource (variation point/variant), deletion of a variant activity, or insertion of a new variant activity. We obtain the following relations:

| Parts | Fields | Values |
|---|---|---|
| Interface | Identification | Variation Point Activity Substitution |
| | Classification | Variation point activity, substitution, configurable process model |
| | Context | {Data substitution (DS), Resource substitution (RS), Variant activity substitution (VAS), variant activity insertion (VAI), variant activity deletion (VAD)} |
| | Problem | Substituting of an activity as a variation point in the configurable process model. |
| | Force | Allows verifying a set of constraints before substituting an activity variation point |
| Realization | Process Solution | **Needed parameters:**<br>*Req_f_VPA: Required functionalities of a variation point activity*<br>*Req_f_OVA: Required functionalities of an old variant activity*<br>*Req_f_NVPA: Required functionalities of a new variation point activity*<br>*F_R: Functionalities of resource*<br>**Design choices :**<br>  a) Substitution by a new variation point activity<br>  b) Substitution of the variation point activity by an existing activity which will be transformed to a variation point<br>**If** the activity is a **new substitute variation point then**<br>  Identify the **old variation point activity** to substitute<br>  Insert type of the **new substitute variation point activity** // *optional or alternative*<br>  Check variants activity compatibility<br>    **If**( Req_f_OVA) belongs to (Req_f_NVPA) **then**<br>      Apply Variant Activity insertion process pattern // *if it is necessary to insert new variants of the new variation point activity*<br>    **Else** Apply Variant Activity deletion process pattern // *to delete variants*<br>    And apply Variant Activity Substitution process pattern // *to substitute uncompatible variants*<br>    **End if**<br>  Apply Data Substitution process pattern<br>  Apply Data Insertion process pattern // *for inserting additional data*<br>  Check required flow sequence condition<br>    **If** incompatible **then**<br>      Substitute (Old_condition,New_condition)<br>    **Else** Insert new condition<br>    **End if**<br>  Apply Resource Substitution process pattern<br>**Else If** we transform an existing activity to a substitute variation point **then**<br>  Repeat the same steps as those of the substitution of a variation point activity by a **new substitute variation point activity**<br>**End if** |

| | Application | **Design choices :** |
|---|---|---|
| | | a) Substitute by the new variation point activity "B" |
| | | b) Substitute the variation point activity by the existing activity "C" which will be transformed to a variation point |
| | | **For (a):** |
| | | 1. We suppose that the designer chooses the variation point activity "A" with A1 and A2 variants to substitute by the new variation point "B" with variant "B1" (1). Insert of type="Null" of "B" (2) |
| | | 2. We have two cases : |
| | | ✓ {R4, R5, R6} of A1 and {R7, R8, R9} of A2 belongs to {R1, R2, R3} of B: Apply variant insertion process pattern to insert B1 and conserve A1 and A2 (2.1) |
| | | ✓ {R4, R5, R6} of A1 and {R7, R8, R9} of A2 does not belongs to {R1, R2, R3} of B : |
| | | • Apply variant activity deletion process pattern to delete A2 (2.2) |
| | | • Apply substitution activity process pattern to substitute A1 by B1 with a new feature =Z and a new configuration constrain= B1 excludes D2. (2.3) |
| | | For example we apply in our case (2.2) and (2.3) |
| | | 3. Apply data substitution process pattern to substitute the old dataobject1 by the new dataobject2 (3) |
| | | 4. Insert new condition=cond2 (4) |
| | | 5. Apply resource substitution process pattern to assign the existing resource to "B" and to insert the variant resource R3 which will perfom "B1" (5) |
| | | **For (b):** |
| | | 1. We suppose that the designer chooses to substitute the variation point activity "A" by the existing activity "C": |
| | | ✓ The sequence flow {A→C} is substituted by the sequence flow {A→D} |
| | | ✓ The "A" is deleted |
| | | 2. Insert type of " C" =Null |
| | | ✓ We suppose that {R4, R5, R6} of A1 and {R7, R8, R9} of A2 belongs to {R1, R2, R3} of C: Apply variant insertion process pattern to insert C1 and conserve A1 and A2 (2) |
| | | 6. Apply data substitution process pattern to substitute the old dataobject1 by the new dataobject2 (3) |
| | | 7. Insert new condition=cond2 (4) |
| | | 8. Apply resource substitution process pattern to assign the existing resource to "C" and to insert the variant resource R3 which will perfom "C1" (5) |
| | | 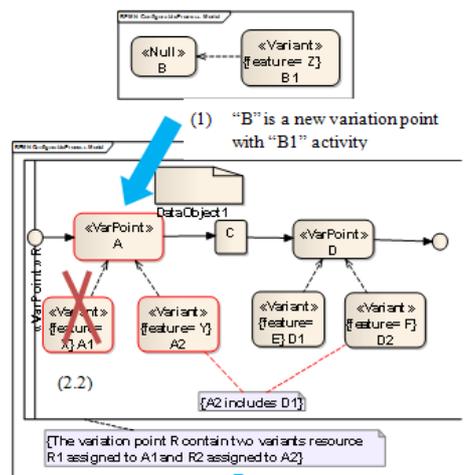 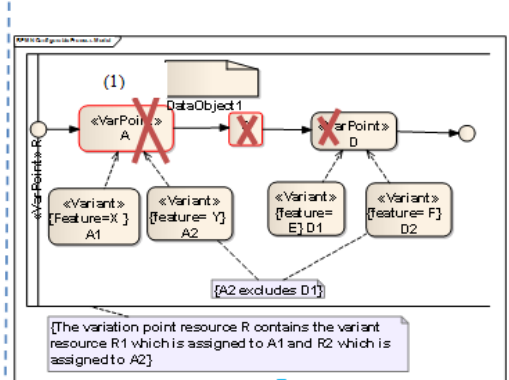 |
| | | *(figure showing two configurable process model diagrams with VarPoint, Variant, DataObject elements and annotations: "{B1 excludes D1}", "{The variation point resource R is updated by adding new variant resource R3 to perform B1}", "{A2 excludes D1}", "{C1 includes D1}", "{The variation point resource R is updated by a new variant resource R3 which is assigned to C1}")* |
| **Relations** | Refines | Activity Substitution (AS) |
| | Uses | {Data Substitution (DS), Resource substitution (RS), Variant Activity Substitution (VAS), Variant Activity Insertion (VAI), Variant Activity deletion (VAD)} |

Table 3 : The VPAS process pattern

We explain below the *Relations* part described in Table 6:
- ➢ {VPAS} **refines** {AS}: the variation point activity substitution process pattern (VPAS) is a specialization of the activity substitution (AS).
- ➢ {VPAS} **uses** {VAS}: to substitute a variation point activity, we have to check if there is a need to substitute related variants or no.
- ➢ {VPAS} **uses** {RS}: to substitute a variation point activity, we have to check if there is a need to substitute the resource or no.
- ➢ {VPAS} **uses** {DS}: to substitute a variation point activity, we have to check if there is a need to substitute data or no.

In this section, we have described in detail process patterns for evolving a configurable process model in terms of activity (variant/variation point). By the same way, we have developed the other process patterns (relative to resource and data) to conduct the different evolution types applied by designers.

## 5. Conclusion and perspectives

In this paper, we have proposed a process patterns system for guiding evolution of configurable process models. Existing approaches for variability management focus on the modeling and configuration of process variants. However, case studies have shown that the evolution of process variants is essential [32]. For this, we have conducted a development methodology to develop a set of reusable process patterns which collaborate in order to model and evolve configurable process models in terms of activity, data and resource. Our proposed process patterns system capitalizes a set of processes and models solutions that perform each step of the evolution process through a reuse process.

We are currently developing a prototype which enables the automation of the proposed approach.